\documentclass[aps,prl,a4paper,twocolumn,showpacs,floatfix,citeautoscript]{revtex4}
\usepackage{graphicx}
\usepackage{amsfonts}
\usepackage{amsmath}
\usepackage{amssymb}
\usepackage{bm}
\usepackage{dcolumn}
\usepackage{hyperref}
\usepackage[usenames]{color}
\hypersetup{pdfborder=0 0 0,colorlinks=true,citecolor=blue,linkcolor=blue}
\begin{document}

\title{Giant switchable Rashba effect in  oxide heterostructures}
\author{Zhicheng Zhong$^1$, Liang Si$^1$, Qinfang Zhang$^2$, Wei-Guo Yin$^3$, Seiji Yunoki$^4$ and Karsten Held$^1$}
\affiliation{$^1$Institute of Solid State Physics, Vienna University of Technology, A-1040 Vienna, Austria \\
$^2$Key Laboratory for Advanced Technology in Environmental Protection of Jiangsu Province, Yancheng Institute of technology, China \\
$^3$Condensed Matter Physics and Materials Science Department, Brookhaven National Laboratory, Upton, New York 11973, USA\\
$^4$Computational Condensed Matter Physics Laboratory, RIKEN, Wako, Saitama 351-I0198, Japan}

\begin{abstract} 
One of the most fundamental phenomena and a reminder of the electron's relativistic nature is the Rashba spin splitting for broken inversion symmetry. Usually this splitting is a tiny relativistic correction,
hardly discernible in experiment. Interfacing a ferroelectric, BaTiO$_3$, and a heavy 5$d$ metal with a large spin-orbit coupling, Ba(Os,Ir)O$_3$, we show 
that giant Rashba spin splittings are indeed possible and even fully controllable by an external electric field.  Based on
density functional theory and a microscopic tight binding understanding, we conclude that the electric field is amplified and stored as a  ferroelectric Ti-O distortion which,  through the network of oxygen octahedra, also induces  a large Os-O distortion.
The BaTiO$_3$/BaOsO$_3$ heterostructure is hence the ideal test station for studying  the fundamentals of the Rashba effect. It allows intriguing application such as the Datta-Das transistor to operate at room temperature.
\end{abstract}
\pacs{73.20.-r, 73.21.-b, 79.60.Jv}
\maketitle

An electric field control of the spin degree of freedom is the key to spintronics and magnetoelectrics \cite{Zutic:rmp04}. In the prototypical
spintronic device, the Datta-Das spin transistor  \cite{Datta:apl90},
an electric field  tunes the Rashba spin splitting
and with that the spin precession frequency. The precession in turn controls the spin polarized current between
two ferromagnetic leads. Microscopically, the Rashba spin splitting originates from the spin orbit coupling (SOC) in a two dimensional electron gas (2DEG) with broken inversion symmetry perpendicular to the 2DEG plane \cite{Rashba:60, Winkler:book}. It has been observed for metal surfaces\cite{LaShell:prl96, Ast:prl07}, semiconductor and oxide heterostructures \cite{Nitta:prl97, Caviglia:prl10b, Shalom:prl10b,Nakamura:prl12, King:prl11, King:prl12}, and even in polar bulk materials \cite{Ishizaka:natm11}.
The Rashba effect splits parabolic bands into two subbands with opposite spin and energy-momentum dispersions  $E^{\pm}({\mathbf k})=(\hbar^{2}{\mathbf k}^{2} / 2 m^{*}) \pm \alpha_{R} |{\mathbf k}|$. Here, $m^{*}$ is the effective mass, ${\mathbf k}$ the  wave vector in the 2DEG plane, and $\alpha_{R}$ the Rashba coefficient which depends on the strength of SOC and inversion asymmetry. An electric field modulates this inversion asymmetry and consequently the Rashba spin splittings. The  electric-field induced change of  $\alpha_{R}$ is  however weak: up to  10$^{-2}$eV\AA \ in semiconductor \cite{Nitta:prl97} or 3$d$ transition metal oxide heterostructures \cite{Caviglia:prl10b, Shalom:prl10b, Nakamura:prl12}.  A large, electric-field tunable Rashba effects  are also  	much sought-after
in the research area of  topological insulators \cite{Bahramy:natcomm12, Nakosai:prl12,Das:natcomm13}.

Giant Rashba effects with $\alpha_R$ of the order of 1eV\AA \ have been reported for metal surfaces \cite{Ast:prl07}, Bi adlayers \cite{Mirhosseini:prb10} and bulk polar materials BiTeI \cite{Ishizaka:natm11,Zhou:screp14}. The large $\alpha_R$ here relies on the surface or interface structural asymmetry, which is hardly changes in an external electric field. To enhance the tunability by an electric field,  Di Sante {\em et al.} \cite{Sante:adv13} hence suggested a ferroelectric semiconductor GeTe. For GeTe, an external electric field  switches between  paraelectric  and ferroelectric phase,  breaking inversion symmetry and tuning on the Rashba spin splitting. At  first glance, this perfectly realizes an electric field control of a giant Rashba effect. However, there is an intrinsic difficulty:  A single material cannot be both, a conductor with large Rashba spin splitting and
a  ferroelectric which necessarily is insulating.

\begin{figure}
\includegraphics[width=\columnwidth]{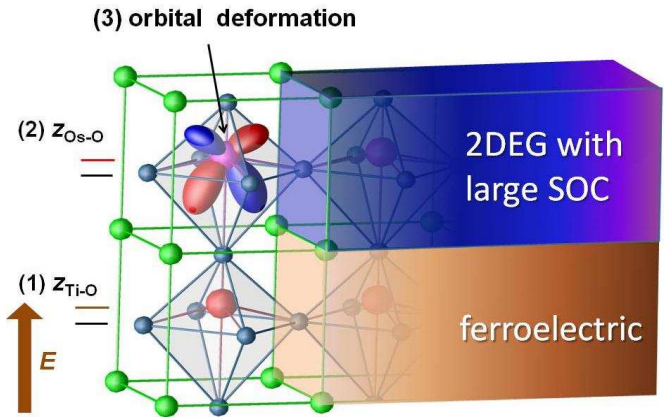}
\caption{Schematics of the mechanism behind the giant switchable Rashba effect:
(1) An external electric field is amplified and stored as a ferroelectric distortion, in particular Ti-O displacements. (2) These displacements entail, via the oxygen octahedron network of the perovskite heterostructure, Os-O displacements. (3) As a consequence of the Os-O displacements the Os orbitals deform, which reflects the broken inversion symmetry. This orbital deformation and the  strong SOC of Os
finally lead to a giant Rashba  spin splittings that is controllable by an electric-field. }
\label{Fig0}
\end{figure}

In this letter, we propose to realize a giant switchable Rashba effect by heterostructures sandwiching a thin metallic film of heavy elements in-between ferroelectric insulators, see Fig.~\ref{Fig0}. The thin film provides a 2DEG with strong SOC, while the inversion asymmetry is induced by  the structural distortion of the ferroelectrics and tunable by an electric field. As a prime example, we study a heterostructure of transition metals oxides, BaOsO$_3$/BaTiO$_3$. BaTiO$_3$ is a well-established ferroelectric \cite{Dawber:rmp05}. It has a simple high-temperature  perovskite structure, with Ba atoms at the edges, a Ti atom at the center and the O atoms at the faces  of a cube. Such a structure has an inversion symmetry center at the Ti site. At room temperature, inversion symmetry is broken since a ferroelectric structural distortion occurs with a sizable Ti-O displacement $z_{\rm Ti-O}$ along one of the the cubic axes. BaOsO$_3$ has been recently synthesized; it is a metallic perovskite with four Os 5$d$  electrons and no sign of magnetism \cite{Shi:jacs13}.  It has a perfect lattice match with BaTiO$_3$. Both materials have the same cation, Ba, which will substantially reduce disorder during epitaxial growth. For BaOsO$_3$/BaTiO$_3$ we find an electric-field tunable Rashba spin splitting which is at least one magnitude larger than the current experimental record \cite{Nitta:prl97, Caviglia:prl10b, Shalom:prl10b,Nakamura:prl12}. The mechanism behind is nontrivial and summarized in Fig.~\ref{Fig0}. Substituting  BaOsO$_3$ by BaIrO$_3$ and  BaRuO$_3$ yields a similar effect; the heterostructure can also be further engineered by varying its thickness and strain.

\begin{figure*}[t!]
\includegraphics[width=7.0 in]{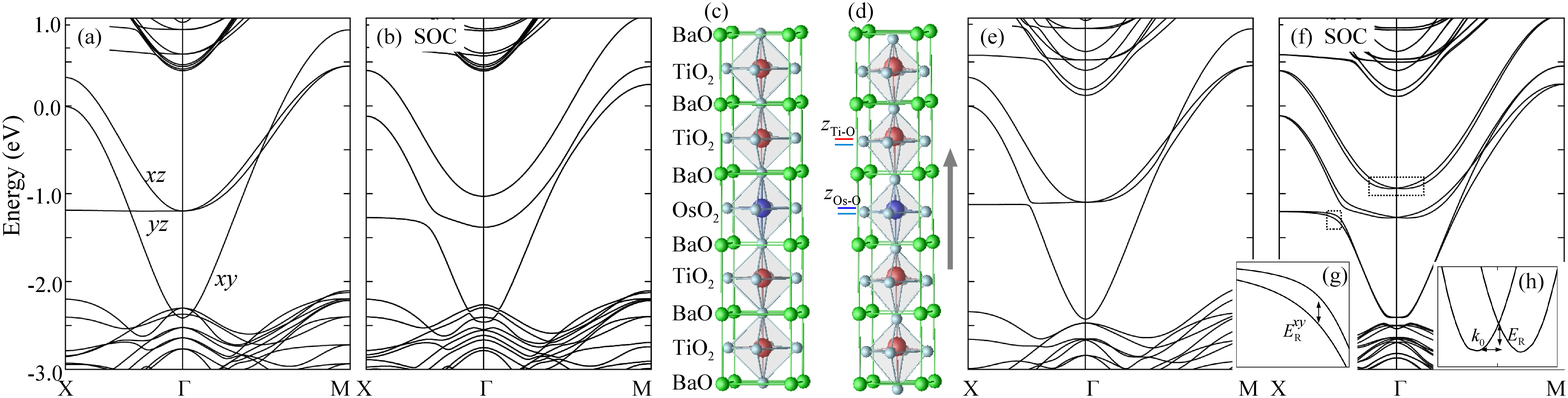}
\caption{DFT calculated band structures  of  (BaOsO$_3$)$_{1}$/(BaTiO$_3$)$_{4}$ multilayers for the paraelectric  (c) and  ferroelectric structure (d).
 The paraelectric heterostructure  breaks cubic but not inversion symmetry
so that the Os $xy$ orbital splits off from two degenerate $yz/xz$ orbitals 
without SOC (a); including SOC further lifts the $yz$/$xz$ degeneracy (b). 
The ferroelectric state further breaks inversion symmetry, which  essentially does not change the bandstructure in the absence of SOC (e) but leads to substantial spin splittings in the presence of SOC (f). The spin splitting $E^{xy}_{\rm R}$ at the $xy-yz$ crossing region is magnified in (g), while the standard Rashba spin splitting $E_{\rm R}$ with momentum offset $k_{0}$ around $\Gamma$ of the upper $yz$/$xz$ mixed orbital is magnified in (h).}
\label{Fig1}
\end{figure*}

{\em Method.} We mainly focus on a  BaOsO$_3$/BaTiO$_3$ heterostructure which has two structural phases: a paraelectric perovskite phase  and a distorted ferroelectric phase, see Fig.~\ref{Fig1}(c) and (d), respectively. We use density-functional theory (DFT) with generalized gradient approximation (GGA) potential \cite{PerdewPRL96} in the Vienna Ab initio Simulation Package (VASP) \cite{Kresse:prb93,Kresse:prb96} and fully relax all the atomic positions, only fixing the paraelectric or ferroelectric symmetries. At zero temperature, the ferroelectric distorted perovskite has the lower GGA energy, but -as in the bulk- the undistorted paraelectric phase will prevail at elevated temperatures.

 Based on the fully relaxed atomic structures, electronic band structures are calculated with modified Becke-Johnson (mBJ) \cite{Tran:prl09} exchange potential as implemented in the Wien2k code \cite{WIEN2k}, which improves the calculated bandgap of BaTiO$_3$. The SOC is included as a perturbation using the scalar-relativistic eigenfunctions of the valence states. Employing  {\em wien2wannier} \cite{Kune20101888}, we project the Wien2k bandstructure   onto maximally localized\cite{Mostofi2008685} Wannier orbitals, from which the orbital deformation is analyzed and a realistic tight binding model is derived \cite{Zhong:prbr13}. We also vary the thickness of the thin films, replace the BaOsO$_3$ by  BaRuO$_3$ or BaIrO$_3$, and simulate the strain effect by fixing the in plane lattice constant to the value of a SrTiO$_3$ substrate. Correlation effects from local Coulomb interaction are studied in the Supplementary Material.

{\em Paraelectric phase.}  We mainly focus on a  BaOsO$_3$/BaTiO$_3$ heterostructure, (BaOsO$_3$)$_{1}$/(BaTiO$_3$)$_{4}$, which consists of one  BaOsO$_3$ layer alternating with four  BaTiO$_3$ layers. Bulk BaTiO$_3$ is a $d^{0}$ insulator: the empty Ti 3$d$ states lie about 3eV above occupied O 2$p$ states. Bulk BaOsO$_3$ is a $d^{4}$ metal: four electrons in the  Os 5$d$  orbitals of $t_{2g}$ character  are 0.8eV above the filled O 2$p$ states. Since BaOsO$_3$ and BaTiO$_3$ share oxygen atoms at the interface, the O 2$p$ states align; and the Os 5$d$ states will stay in the energy gap of BaTiO$_3$. 
Hence, the density functional theory (DFT) calculated band structures of BaOsO$_3$/BaTiO$_3$ in Fig.~\ref{Fig1} shows three Os $t_{2g}$ ($xy$, $yz$, $xz$) bands near the Fermi level;
they are  dispersionless along the $z$ direction (not shown).
That is, the low energy electronic degrees of freedom are  confined by  the
insulating BaTiO$_3$ layers to the OsO$_2$ plane, forming a 2DEG. Already in the high temperature paraelectric phase, 
 this heterostructure confinement reduces the initial cubic symmetry of the Os $t_{2g}$ orbitals in the bulk perovskite: At $\Gamma$ the $xy$ band in Fig.~\ref{Fig1}~(a) is 1.2eV lower in energy than the degenerate $yz$/$xz$ bands (Fig.~\ref{Fig1}(a)). Including the SOC, the $yz$/$xz$ doublet splits into two subbands with mixed $yz$/$xz$ orbital character, see Fig.~\ref{Fig1}(b). For the paraelectric phase of Fig.~\ref{Fig1}(a-c), the spin degeneracy is however still preserved.

\begin{figure}[!]
\includegraphics[width=0.9 \columnwidth]{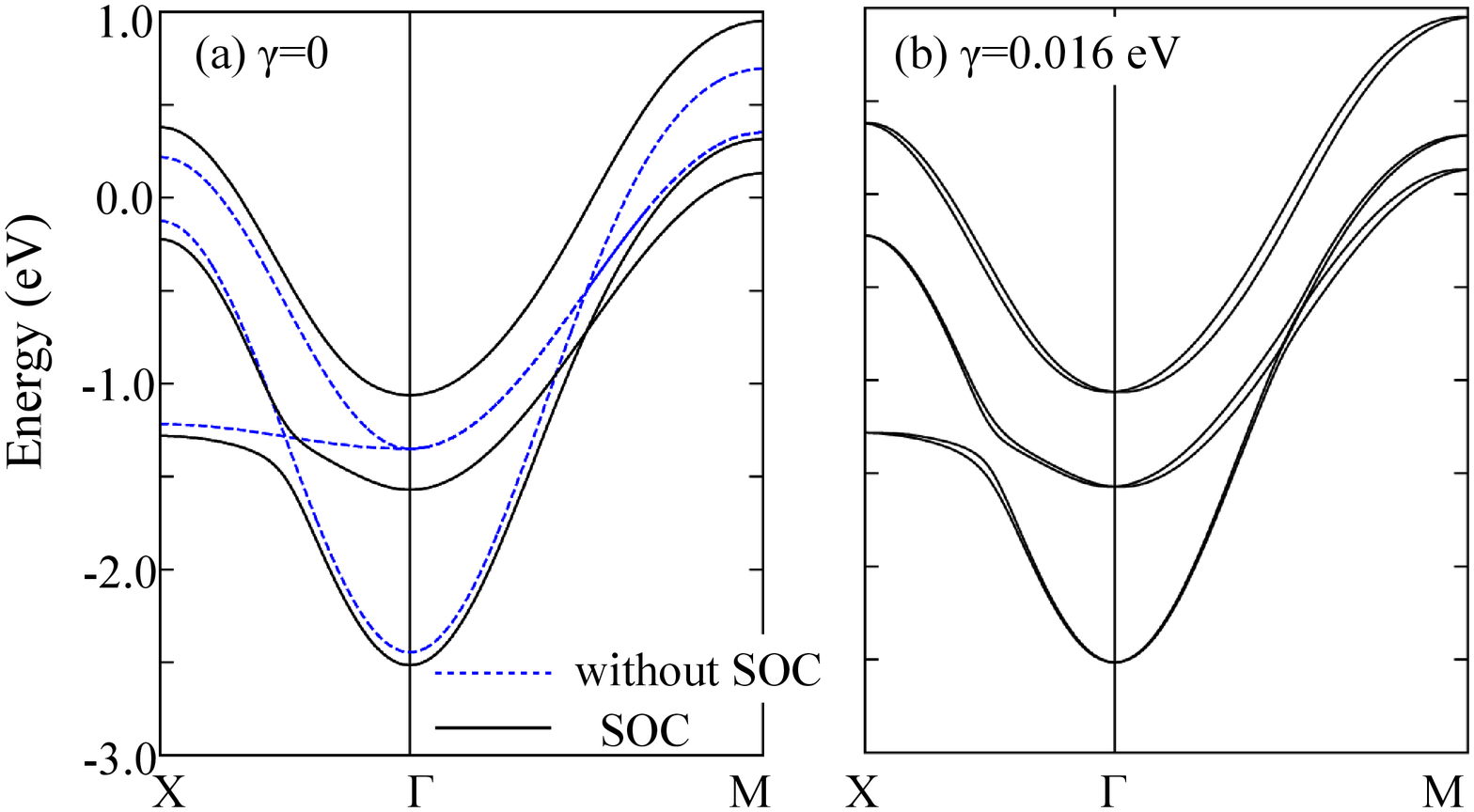}
\caption{Tight binding bandstructure of the three Os $t_{2g}$ orbitals
for (a) the paraelectric case with and without SOC
and (b) the ferroelectric case with SOC and asymmetry parameter $\gamma=0.016\,$eV taken from the Wannier projection.}
\label{Fig2}
\end{figure}

To better understand the DFT results, we construct an interface hopping Hamiltonian $H_{0}^{i}$ \cite{Zhong:prb13}  for the 2DEG confined Os $t_{2g}$ electrons
by a Wannier projection. The energy-momentum  dispersion for the $xy$ orbital is $\epsilon({\mathbf k})^{xy}=-2t_{1}\cos k_{x}-2t_{1}\cos k_{y}-4t_{3}\cos k_{x}\cos k_{y}$, while that of the  $yz$   is $\epsilon({\mathbf k})^{yz}=-2t_{2}\cos k_{x}-2t_{1}\cos k_{y}$  ($\epsilon({\mathbf k})^{xz}$ is symmetrically related by $x \leftrightarrow y$). The largest hopping $t_{1}=0.392$eV is along the
direction(s) of the orbital lobes, $t_{2}=0.033$eV and $t_{3}=0.094$eV indicate a much smaller hopping perpendicular to the lobes and along $(1, 1, 0)$,  respectively. The $xy$--$yz/xz$ orbital splitting  at $\Gamma$ is $2t_{1}-2t_{2}+4t_{3}=1.1$eV  and arises from the anisotropy of the orbitals. Overall, the three parameter tight binding model  in Fig.~\ref{Fig2}(a) (dashed line) is consistent  with  DFT in
Fig.~\ref{Fig1}(a). Next, we include the atomic SOC Hamiltonian  $H_{\xi}=\xi \vec{l}\cdot \vec{s}$, expressed in the $t_{2g}$ basis with $\xi=0.44$eV for Os. It  lifts the $yz/xz$ degeneracy and yields good agreement with DFT, cf.\ Fig.~\ref{Fig2}(a)  (solid line) and  Fig.~\ref{Fig1}(b). 


{\em Ferroelectric phase.}
In sharp contrast to the paraelectric case, ferroelectrically distorted BaOsO$_3$/BaTiO$_3$ exhibits evident spin splittings in the presence of SOC, see Fig.~\ref{Fig1}(f). Note, in the absence of SOC the DFT band structure in Fig.~\ref{Fig1}(e) is still very similar to the paraelectric case in Fig.~\ref{Fig1}(a). The spin splitting $E^{xy}_{\rm R}$ of  the $xy$ band is small around $\Gamma$, but strongly enhanced up to 50meV around the $xy$/$yz$ crossing region as shown in Fig.~\ref{Fig1}(g), cf.\   Table~\ref{alldata}, . The other two subbands of $yz$/$xz$ character also show a strong splitting around $\Gamma$. In this respect,
the upper  $yz$/$xz$ subband exhibits a standard Rashba behavior, see the magnification in Fig.~\ref{Fig1}(h), where the momentum offset $k_{0}=0.043\,$\AA$^{-1}$, the  Rashba energy $E_{\rm R}=4\,$meV, and hence $\alpha_{\rm R}=2E_{\rm R}/k_{0} = 0.186\,$eV\AA. The  behavior of the lower  $yz$/$xz$ subband  is more complex (distorted) due to the proximity of the crossing region.

These spin splittings arise from ferroelectric distortions which break the inversion symmetry. Bulk BaOsO$_3$ has no such ferroelectric distortions 
and is inversion symmetric with  Os and O atom  in the same plane.
Also for the paraelectric heterostructure, the OsO$_2$ layer is still an  inversion plane. The ferroelectric heterostructure breaks inversion symmetry. As listed in Table \ref{alldata}, the averaged ferroelectric Ti-O displacement in BaTiO$_3$ layers is around 0.14 \AA, which efficiently induces a sizable Os-O displacement of 0.05~\AA \ via the oxygen octahedron network of the perovskite structure. This structural distortion modifies the crystal environment of Os $t_{2g}$ orbitals and deforms the orbital lobes, see Fig.~\ref{Fig0}.

In order to quantify the orbital deformation, we project the DFT results above onto maximally localized Wannier orbitals and directly extract the key parameter: a directional, spin-independent inter-orbital hopping  $\gamma=\langle xy|H|yz(R)\rangle$, were $R$ is the nearest neighbor in $x$ direction. In ${\mathbf k}$-space one gets as a matrix for the $yz$,$xz$,$xy$  orbitals (independent of spin)
\cite{Zhong:prbr13}
\begin{equation*} 
H_{\gamma}= \gamma\left(\begin{array}{cccccc}
0   & 0  & 2i\sin k_{x}\\
0  & 0  & 2i\sin k_{y}\\
-2i\sin k_{x}  & -2i\sin k_{y}  & 0\end{array}\right) \; .
\end{equation*}.
 We obtain $\gamma= 0.016\,$eV for the ferroelectric 
heterostructure, while for the paraelectric case  $\gamma=0$ due to inversion symmetry.  Let us emphasize that  $\gamma$ is the relevant
 measure for the orbital deformation and 
inversion symmetry breaking.  In combination with the strong SOC of OS 5$d$ electrons,  $H_{\gamma}$ results in the present giant Rashba  spin splitting. 

The  Hamiltonian   $H_{0}^{i}+H_{\xi}+H_{\gamma}$
already well describes the spin splittings of the $xy$ orbital in good agreement with DFT results, but underestimates the spin splittings of the $yz$/$xz$ subbands near $\Gamma$. The reason for this is that the atomic  SOC is not an accurate description anymore because the  SOC of Os is too strong and the deviation
from the inversion symmetry too large. Hence, we need to go beyond the purely 
atomic SOC and include an inversion asymmetry correction to
the  SOC matrix so that $H_{\xi}$ in the $t_{2g}$ basis  ($yz|\!\uparrow\rangle$, $yz|\!\downarrow\rangle$, $xz|\!\uparrow\rangle$, $xz|\!\downarrow\rangle$, $xy|\!\uparrow\rangle$, $xy|\!\downarrow\rangle$) reads

\begin{equation*}
 \left(\begin{array}{cccccc}
0 & 2 \gamma ' \sin k_{y} & i\frac{\xi}{2} & 0 & 0 & -\frac{\xi}{2}\\
2 \gamma ' \sin k_{y} & 0 & 0 & -i \frac{\xi}{2}& \frac{\xi}{2} & 0\\
-i\frac{\xi}{2} & 0 & 0 & -2 \gamma ' \sin k_{x} & 0 & i\frac{\xi}{2}\\
0 & i\frac{\xi}{2} & -2 \gamma ' \sin k_{x} & 0 & i \frac{\xi}{2}& 0\\
0 & \frac{\xi}{2} & 0 & -i\frac{\xi}{2} & 0 & 0\\
-\frac{\xi}{2} & 0 & -i\frac{\xi}{2} & 0 & 0 & 0
\end{array}\right).
\end{equation*}
For  the bandstructure of Fig.~\ref{Fig2}(b) we have taken $\gamma '=0.022$ eV
from the Wannier projection, yielding good agreement with the DFT results Fig.~\ref{Fig1}(f).

\begin{table*}[!]
\begin{ruledtabular}
\caption{Polar distortions and Rashba spin splittings of (BaOsO$_3$)$_{n}$/(BaTiO$_3$)$_{m}$ multilayers, alternating $n$ layers of BaOsO$_3$ and $m$ layers of BaTiO$_3$. Second and third column: averaged displacement of Os-O and Ti-O along $z$ as calculated by DFT. Forth column:  strength of the interface asymmetry term $\gamma$ obtained from the Wannier projection. Fifth to seventh column: DFT calculated parameters characterizing Rashba effects: momentum offset $k_{0}$, Rashba energy $E_{\rm R}$, and Rashba coefficient  $\alpha_{\rm R}$ of the upper $yz$/$xz$ subband around $\Gamma$ as defined in Fig.~\ref{Fig1}(h). Eighth column: spin splitting $E^{xy}_{\rm R}$ around the crossing region of $xy$ and $yz$ orbitals as in Fig. 1(g). To demonstrate the effects of heterostructure engineering, we also list the results of  BaOsO$_3$/BaTiO$_3$ strained by a SrTiO$_3$ substrate, as well as BaRuO$_3$/BaTiO$_3$ and BaIrO$_3$/BaTiO$_3$. Band structures are given in Supplementary Material.}
\begin{tabular}{l l l l l l l l l l l l l l l l}
$n:m$ & Os-O (\AA) & Ti-O (\AA) &  $\gamma$ (eV) & $k_{0}$ (\AA$^{-1}$) & $E_{\rm R}$ (eV) & $\alpha_{\rm R}$ (eV\AA) & $E^{xy}_{\rm R}$ (eV)  \\
\hline
1:3 BaOsO$_3$/BaTiO$_3$ & 0.038 & 0.065 & 0.011 & 0.025 & 0.001 & 0.08 & 0.031 \\
1:4 & 0.049 & 0.103 & 0.016 & 0.043 & 0.004 & 0.186 & 0.053 \\
2:3 & $<$0.010 & $<$0.010 & 0.003 & 0 & 0 & - & 0.004 \\
2:4 & 0.024 & 0.071 & 0.010 & 0.035 & 0.002 & 0.114 & 0.018 \\
1:3 strained & 0.095 & 0.223 & 0.030 & 0.105 & 0.021 & 0.396 & 0.102 \\
1:4 strained & 0.099 & 0.229 & 0.031 & 0.115 & 0.024 & 0.417 & 0.106 \\
\hline
1:4 BaIrO$_3$/BaTiO$_3$ & 0.115& 0.119 & 0.021 & 0.145 & 0.053 & 0.731 & 0.051 \\
1:4 BaRuO$_3$/BaTiO$_3$ & 0.082 & 0.119 & 0.015 & 0.128 & 0.016 & 0.250 & 0.007 \\
\end{tabular}
\label{alldata}
\end{ruledtabular}
\end{table*}

{\em Electric-field tunability.} The Rashba spin splittings can be tuned by an electric field because of the ferroelectricity of BaTiO$_3$. It is a nature of ferroelectrics that an external electric field can  tune and even reverse the polarization and structural distortion \cite{Mathews:sc97, Dawber:rmp05,Duan:prl06, Mirhosseini:prb10, Tazaki:jpcm09}, including that of the BaOsO$_3$ layer. 
Because of the BaOsO$_3$ layer, the ferroelectricity of the heterostructure is 
not as strong as in   bulk BaTiO$_3$. Hence
 a smaller electric field change than in the bulk (E=10$^{3}$V/cm \cite{Tazaki:jpcm09})
is be needed for going through the ferroelectric hysteresis loop. Such an electric field can tune the  Rashba coefficient
listed in   Table \ref{alldata}, which are at least one order of magnitude  more larger than in semiconductors \cite{Nitta:prl97} or 3$d$ oxide heterostructures \cite{Caviglia:prl10b, Shalom:prl10b,Nakamura:prl12}.
For applications it is of particular importance that $E_R$ now clearly exceeds 
$0.025\,$eV, the thermal energy at room temperature. 
 Furthermore, we propose to take advantage of the ferroelectric hysteresis and the remnant  ferroelectric polarization,  using our heterostructure  as a spintronics memory device. 

{\em Heterostructure engineering.} Besides applying an electric field, heterostructure engineering such as varying its thickness, strain, and the material combination is a powerful way to tune the Rashba spin splittings. First, we vary the thickness of BaOsO$_3$ and BaTiO$_3$. It is well known experimentally and theoretically \cite{Junquera:nat03,Dawber:rmp05} that below a few nanometer  thickness of the BaTiO$_3$  film, the Ti-O displacement is reduced and eventually ferroelectricity vanishes. As shown in Table \ref{alldata}, when decreasing the thickness of BaTiO$_3$ and increasing that of BaOsO$_3$, the Rashba spin splittings is indeed substantially reduced. Second, ferroelectricity of BaTiO$_3$ is sensitive to strain \cite{Choi:sc04}. Taking SrTiO$_3$ as a substrate provides a 2.5\% in-plane compressive strain. This enhances the ferroelectric distortion as well as Rashba spin splitting, $\alpha_{\rm R}$=0.417eV\AA \ in Table \ref{alldata}. Given the giant Rashba splitting, photoemission and transport measurements  should be able to validate this effect by comparing thin films with different thickness and substrate. Third, we replace Os by other 5$d$ elements for shifting the Fermi energy; and also since some Os oxides are toxic. Indeed, similar physics can be found in BaIrO$_3$/BaTiO$_3$ with $\alpha_{\rm R}=0.731eV\,$\AA, see Table \ref{alldata}, and in  BaRuO$_3$/BaTiO$_3$. The latter also offers a platform to study the interplay of Rashba physics and electronic correlations; for instance, novel phenomena is expected in mixed 3$d$-5$d$ transition-metal materials \cite{Yin:prl13}. Note a strong temperature dependence of the $\alpha_R$ in SrIrO$_3$ was recently reported \cite{Zhang:arXiv13}.

{\em Conclusion.} We propose to combine  a heavy 5$d$ compound such as BaOsO$_3$ and a ferroelectric such as BaTiO$_3$ in a heterostructure for realizing giant Rashba spin splittings that are switchable by an electric field. The physics behind is that ferroelectric  BaTiO$_3$ amplifies the electric field through a distortion (polarization)
of its   TiO$_6$ octahedra. This distortion very efficiently  propagates to the OsO$_6$ octahedron via the shared oxygens,  breaking inversion symmetry and
deforming the Os orbital lobes. Together with the strong SOC 
it leads to a giant  tunable Rashba effect with, e.g., $\alpha_{\rm R}$=0.731eV\AA \ for BaIrO$_3$/BaTiO$_3$. This is at least one  magnitude larger than state-of-the-art. Moreover, the ferroelectric  hysteresis  and remnant polarization leads to a memory effect which can be exploited for spintronics memory devices. 

{\em Acknowledgments.} ZZ acknowledges financial support by the Austrian Science Fund through the SFB ViCoM F4103-N13, QFZ by NSFC (11204265), the NSF of Jiangsu Province (BK2012248), KH by the European Research Council under the European Union's Seventh Framework Program (FP/2007-2013)/ERC through grant agreement n.\ 306447, and WY by the U.S. Department of Energy under Contract No. DE-AC02-98CH10886. Calculations have been done on the Vienna Scientific Cluster~(VSC).


\end{document}